# A Proactive Insider Threat Management Framework Using Explainable Machine Learning

**Selma Shikonde**
u24783902@tuks.co.za
**Mike Wa Nkongolo**
mike.wankongolo@up.ac.za

Department of Informatics, University of Pretoria, South Africa

## Abstract

**Problem Statement.** Over the years, the technological landscape has evolved, reshaping the overall security posture of organisations. This evolution has increased exposure to cybersecurity threats, many of which originate from within the organisation itself. Insider threats continue to pose significant challenges to institutions across various sectors, particularly where cybersecurity infrastructure, expertise, and regulatory mechanisms are still developing.

**Aims.** The objective of this study is to propose a framework and strategies for proactively managing insider threats through the implementation of robust security and privacy measures. The proposed framework, termed the Insider Threat Explainable Machine Learning (IT-XML) framework, integrates the Cross-Industry Standard Process for Data Mining (CRISP-DM) methodology with Hidden Markov Models (HMM) to enhance proactive insider threat management and decision-making.

**Methods.** A quantitative research approach is employed, utilising a structured online questionnaire consisting of five sections that assess employees' knowledge of insider threat patterns and detection, access control and privacy levels, existing security measures, policy gaps, and proactive mechanisms in place. Participants were drawn from three large organisations operating within a data-sensitive environment. The IT-XML framework addresses critical gaps in current insider threat management approaches by providing organisational assessment capabilities through survey-based data collection, HMM-driven pattern recognition for security maturity classification, and evidence-based recommendations for proactive threat mitigation.

**Contribution.** From a scientific standpoint, the study contributes to the cybersecurity and data mining domains by developing a dataset on insider threat management practices and demonstrating how the integration of CRISP-DM and HMM can improve insider threat modelling and prediction.

**Results.** The framework successfully classified all participating organisations at the developing security maturity level, achieving 97–98% confidence using HMM. This finding highlights the potential for organisations to collaboratively enhance their security maturity through shared assessment and benchmarking. Empirical results indicated that information-sharing violations were the most common insider threat (61.7%), followed by unauthorised access incidents (46.7%). The framework achieved a classification accuracy of 91.7%, identifying audit log access limits as the most critical security control. Further evaluation using Random Forest (RF) revealed that vendor breach notification requirements (0.081) and regular audit log reviews (0.052) were key determinants of insider threat resilience. Additional explainability analyses using SHAP and LIME enhanced the transparency and interpretability of the framework's predictions.

**Relevance.** In practice, the research promotes a collaborative security enhancement approach based on shared maturity profiles, encouraging organisations to move from isolated practices to collective capacity-building initiatives. Implementing the IT-XML framework enables organisations to strengthen their insider threat management capabilities and enhance overall security resilience—ensuring better protection of critical information assets.

**Keywords:** Insider Threat Management, Explainable Machine Learning, Hidden Markov Models, Cybersecurity Framework

## Introduction

*Insider Threats in the Digital Era*

The rapid expansion of the digital world and the widespread adoption of online transactions have increased the risk of cybercrime (Kuzior et al., 2024). As cybersecurity threats escalate globally (Kuzior, 2024), organisations are compelled to strengthen mechanisms for protecting sensitive data and information systems against both external and internal attacks. Among these, insider threats have emerged as one of the most critical challenges, contributing substantially to data leakage and security breaches (Shaikh et al., 2023). Insider threats can be malicious or non-malicious, encompassing acts such as intellectual property theft, sabotage, negligence, and unauthorised disclosure of confidential information (Allen et al., 2024). Their impact is amplified because insiders often possess elevated privileges or direct access to critical organisational assets. According to Yuan and Wu (2021), insider threats are among the most complex and damaging forms of cyber risk, often resulting in severe consequences for data integrity and organisational resilience.

Kuzior et al. (2024) analysed global cybercrime trends between 2016 and 2023 across 33 countries, revealing that cybercrime incidents grew by more than 50% in nations such as Malta, Slovenia, Iceland, Moldova, and Slovakia. Their study highlighted striking regional variations, with Greece, Belgium, France, and Germany recording the highest cybercrime rates, while Moldova, Georgia, Slovenia, and Iceland reported relatively lower levels. Similarly, Richards (2021) noted an increase in cybercrime within African governments, primarily due to insufficient legal frameworks and inadequate policy enforcement. These findings underline the vulnerability of public sector organisations, which often face challenges such as weak governance structures, limited policy integration, and slower implementation of security measures compared to private entities. Such constraints complicate efforts to protect national infrastructure and essential public services from cascading cyber impacts. To mitigate insider threats effectively, organisations must simplify and enforce security policies, enhance employee awareness, and foster a culture of cybersecurity accountability. A study by PwC (2023) found that organisations with comprehensive security policies and regular staff training experience up to 60% fewer insider incidents than those lacking structured awareness programs. Figure 1 illustrates key insider threat factors observed within large organisations.

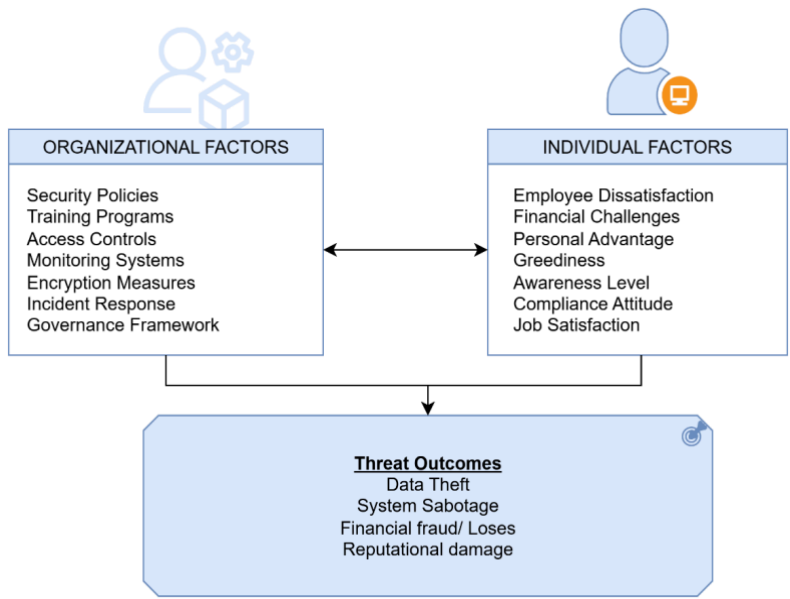

**Figure 1.** Organisational Insider Factors

Figure 1 illustrates the interaction between organisational and individual factors that contribute to diverse insider threat outcomes. Organisational factors highlight the necessity of implementing robust access controls, encryption, and proactive security measures to safeguard critical assets (Srivastava et al., 2024; Schütz et al., 2023; Silaule et al., 2022). Conversely, individual factors emphasise the motivational drivers behind insider threats, including employee dissatisfaction, greed, personal gain, and financial distress (Whitelaw et al., 2024; Saxena, 2020). The resulting threats encompass data theft, data breaches, fraud (Saxena, 2020), financial losses, reputational harm (Kuzior, 2024; Maheshwary et al., 2024), and unauthorised access. Ophoff et al. (2014) discussed the importance of examining insider threats through governance and compliance frameworks, while Bore (2020) stressed the need to monitor and control privileged user access to ensure security and mitigate risks to organisational processes. Accordingly, organisations should invest in advanced security mechanisms such as intrusion detection and prevention systems, encryption protocols, and firewalls to protect cloud-stored data (Deep & Jain, 2023). Given the increasing frequency of privacy and data breaches, implementing comprehensive data protection measures remains a critical priority (Silaule et al., 2022).

**Insider Threat Challenges**

*State-Owned Enterprises (SOEs)*

SOEs face unique insider threat challenges distinct from those of private sector organisations due to their complex governance structures, regulatory compliance obligations, and heightened public accountability. Unlike private companies, SOEs must navigate political oversight, manage critical infrastructure, and protect sensitive governmental data. Their high degree of state ownership often leads to coordination difficulties and ambiguous roles, making insider threat management particularly challenging (Bruton et al., 2015). The increasing prevalence of cybersecurity threats, particularly insider threats, continues to challenge organisations in protecting their information assets (Silaule et al., 2022). Insider threats—whether intentional or unintentional—undermine organisational security policies through malicious actions or negligent behaviours. The absence of effective insider threat management strategies has resulted in growing incidents of data loss and privilege misuse (Whitelaw et al., 2024).

In the African context, insider threat concerns are further amplified by limited resources and a shortage of cybersecurity expertise. Ndjavera (2023) reported a 40% increase in cyberattacks, while Nangolo (2023) highlighted that a lack of cybersecurity awareness and skilled personnel remains a significant barrier. These constraints position cybersecurity as a reactive rather than proactive practice in many African SOEs. Allen et al. (2024) argue that many African organisations still lack efficient mechanisms for detecting and mitigating insider threats, leaving information security at considerable risk (Hong et al., 2023). The consequences include data breaches (Abuasal et al., 2024; Shaikh et al., 2023), fraud and unauthorised access (Renaud et al., 2024), financial losses, reputational damage (Legg et al., 2017; Ophoff et al., 2014), legal implications (Milson & Demir, 2023), and theft of intellectual property (Shaikh et al., 2023).

Existing insider threat management approaches, largely designed for private sector settings, fail to address the specific operational realities of African SOEs. These include regulatory compliance, political oversight, and public governance expectations. In developing economies such as South Africa, SOEs play a critical role in national infrastructure but often lack sufficient cybersecurity frameworks, funding, and skilled personnel to combat evolving threats effectively. To address these gaps, insider threat management in SOEs requires a holistic approach encompassing robust access controls, encryption (Srivastava et al., 2024), proactive monitoring (Schütz et al., 2023a; Silaule et al., 2022), and a strong security culture (Milson & Demir, 2023). Further research is needed to establish ethical frameworks guiding cybersecurity practices and vulnerability management (Nygård & Katsikas, 2024). The absence of effective detection and mitigation mechanisms continues to expose organisations to significant risks (Allen et al., 2024). Mohd and Yunos (2020) further caution that without comprehensive strategies, insider threats can severely compromise organisational performance, underscoring the importance of Data Leak Prevention (DLP) mechanisms.

**Research Focus**

This study responds to the critical need for a SOE-specific insider threat management framework, termed the IT-XML Framework, integrating the CRISP-DM process with Hidden Markov Model (HMM) methodology. The CRISP-DM (Cross Industry Standard Process for Data Mining) methodology provides a structured, six-stage approach—business understanding, data understanding, data preparation, modelling, evaluation, and deployment—guiding the study from the contextual analysis of SOEs to framework development. This ensures a systematic examination of insider threat dynamics through survey-based data analysis. The HMM component applies a Machine Learning (ML) approach to identify hidden behavioural patterns and classify organisations into basic, developing, or advanced security maturity levels. This dual-method design enables the discovery of patterns in organisational behaviour and security practices, thereby supporting evidence-based strategies for proactive insider threat management.

## Research Objectives

The aim of this research is to develop a framework and strategies for proactive insider threat management within SOEs through robust security and privacy measures, supported by a dedicated dataset.

Objective 1: Develop a comprehensive dataset on insider threat patterns within SOEs and construct a taxonomic classification system for threat types specific to SOE environments.
 Objective 2: Analyse how authorised access levels contribute to privacy vulnerabilities in SOE operational contexts using survey data integrated with HMM analysis.
 Objective 3: Design and validate a security maturity assessment tool that produces actionable recommendations to strengthen organisational security posture and identify implementation gaps.
 Objective 4: Develop the IT-XML Framework by integrating CRISP-DM methodology with HMM to generate evidence-based strategies and recommendations for proactive insider threat management in SOEs.

These objectives are achieved through a systematic, survey-based assessment approach that facilitates comprehensive data collection across multiple SOE environments. The methodology supports the CRISP-DM analytical process while providing the behavioural insights necessary for HMM modelling, ultimately contributing to a deeper understanding of how organisational characteristics, policy effectiveness, and employee behaviour interact to create insider threat vulnerabilities in SOEs.

## Research Questions

The research questions are aligned with the study's comprehensive six-section survey framework designed specifically for SOE environments. These questions address descriptive, diagnostic, and prescriptive dimensions—focusing on threat identification, privacy implications, and proactive mitigation strategies.

**Primary Research Question:**
 *In what ways can SOEs proactively manage insider threats while ensuring the privacy and security of critical information?*

**Sub-Research Questions:**

1. *What are the main types and patterns of insider threats identified in SOEs based on the constructed dataset?*

2. *How does authorised access to sensitive data affect the overall privacy landscape within SOEs?*

3. *What security measures are currently implemented in SOEs to mitigate insider threats?*

4. *What gaps exist in existing security policies and procedures, and how can they be addressed?*

5. *What proactive measures can SOEs adopt to strengthen their capacity for insider threat management?*

**Assumptions**

The assumptions underlying this research on insider threat management, privacy, and security in SOEs include the following:

- Participant Integrity: It is assumed that SOE employees and stakeholders will voluntarily participate in the survey and provide honest and accurate responses regarding organisational security practices and insider threat experiences.

- Representative Diversity: The selected SOEs represent adequate diversity in size, sector, and operational complexity, allowing meaningful generalisation of the findings to the broader SOE context.

- Methodological Validity: A survey-based assessment approach can effectively capture organisational security maturity levels and insider threat management capabilities through structured data collection.

- Pattern Variability: Current insider threat management practices within SOEs exhibit sufficient variation to enable meaningful pattern recognition and classification using HMM analysis.

These assumptions are essential for the research to effectively address the challenges faced by SOEs in managing insider threats while maintaining a balance between security, privacy, and organisational efficiency.

**Research Contributions**

This research provides both scientific and practical contributions to the body of knowledge on insider threat management. The key contributions are summarised below.

Scientific Contributions

a) The study contributes scientifically through the development of the proposed Insider Threat Explainable Machine Learning (IT-XML) framework, which integrates the Cross-Industry Standard Process for Data Mining (CRISP-DM) methodology with Hidden Markov Models (HMM) to address insider threat management within public sector organisations.

b) It fills existing research gaps by creating a comprehensive insider threat dataset specific to the public enterprise environment, serving as a foundation for future research on data-driven threat detection and management strategies.

c) The research expands the theoretical understanding of how CRISP-DM and HMM methodologies can be effectively combined to enhance the detection, classification, and prediction of insider threats.

d) By applying these methodologies in the context of insider threat analysis, the study provides a scientific advancement to the fields of cybersecurity, data mining, and organisational risk management, enriching both theoretical and empirical perspectives in these domains.

Practical Contributions

a) The study unveils unique challenges and opportunities in insider threat management within public organisations, providing a practical framework that can be adapted across diverse institutional contexts.

b) Implementation of the IT-XML framework enables organisations to strengthen their capacity to manage insider threats through improved security assessment, monitoring, and policy alignment, thereby ensuring better protection of critical information assets.

c) The framework allows management teams to perform systematic and standardised security evaluations using survey-based data, enabling benchmarking against security maturity models and facilitating evidence-based investment decisions.

d) The insider threat dataset developed in this study provides organisational leaders with data-driven insights into threat patterns and vulnerabilities, supporting informed decision-making related to resource allocation, policy formulation, and staff training initiatives.

e) The survey-based assessment tool offers a practical mechanism for conducting regular security audits and tracking improvement over time, helping organisations to measure and demonstrate the effectiveness of security investments to internal and external stakeholders.

f) Finally, the study translates its findings into actionable recommendations that organisational leaders can implement immediately, enabling a shift from reactive incident response to proactive threat prevention and the promotion of a sustainable security culture within their institutions.

**Literature Review**

The literature review followed a structured search and selection process guided by the PRISMA (Preferred Reporting Items for Systematic Reviews and Meta-Analyses) guidelines to ensure systematicity, transparency, and reproducibility in identifying and selecting relevant sources (Page et al., 2021).

Search Method

A comprehensive search was conducted between January 2019 and September 2025 using multiple reputable databases and digital libraries, including the University of Pretoria's (UP) Computer Science Database, Google Scholar, IEEE Xplore, ACM Digital Library, and cybersecurity institutional repositories. Additional relevant online publications were also consulted.

The search strategy employed a combination of keywords and Boolean operators such as:

*"insider threat\*" OR "internal threat\*" AND "cybersecurity" OR "information security" AND "state-owned enterprise\*" OR "SOE\*" OR "public sector" AND "CRISP-DM" OR "Hidden Markov Model" OR "Random Forest" OR "machine learning" OR "SHAP" OR "LIME".*

These terms were carefully tailored to reflect the state-owned enterprise (SOE) context and the methodological frameworks relevant to this study.

Inclusion Criteria

Sources were included if they met the following criteria:

1. Focused on insider threat management, detection, or prevention.

2. Addressed cybersecurity frameworks applicable to SOEs or public sector organisations.

3. Employed methodologies such as CRISP-DM, Hidden Markov Models (HMM), Random Forest (RF), or explainable AI techniques (SHAP/LIME).

4. Were peer-reviewed with clearly defined methodologies.

5. Published in English between 2019 and 2025.

Exclusion Criteria

Studies were excluded if they:

1. Lacked a clear methodology or empirical validation.

2. Focused exclusively on external threats without insider threat components.

3. Were opinion pieces, non-peer-reviewed blogs, or informal reports.

4. Represented duplicate publications.

5. Lacked organisational context relevant to SOEs.

Selection Process

The initial database search identified 155 records. After removing 22 duplicates, 133 records were screened based on title and abstract, leading to the exclusion of 78 that were deemed irrelevant. The remaining 55 articles underwent full-text assessment, resulting in the exclusion of 17 studies due to reasons such as:

- Lack of SOE relevance (n=6)
- Absence of insider threat focus (n=5)
- Methodological limitations (n=3)
- Geographic mismatch (n=2)
- Inaccessibility (n=1)

Consequently, 38 studies were included in the final review. Figure 2 presents the PRISMA flow diagram, illustrating the systematic literature selection process for the Insider Threat Explainable Machine Learning (IT-XML) framework development. Figure 2 shows the selection process using PRISMA commencing with 155 sources identified during initial searches, followed by elimination of irrelevant sources and nonacademic related articles. Thereafter, only 38 articles were utilised for the detailed review due to their relevance to insider threat management and their applicability to SOE environments. This approach assisted with focused research on literature that contributed to the development of the IT-XML framework for SOEs. There are 38 included studies which are categorised in five areas relevant to insider threat management in SOEs. These studies were published between 2014 and 2025 with the majority (n=28, 74%) published since 2020 highlighting the growing research interest in insider threat detection and organisational cybersecurity maturity.

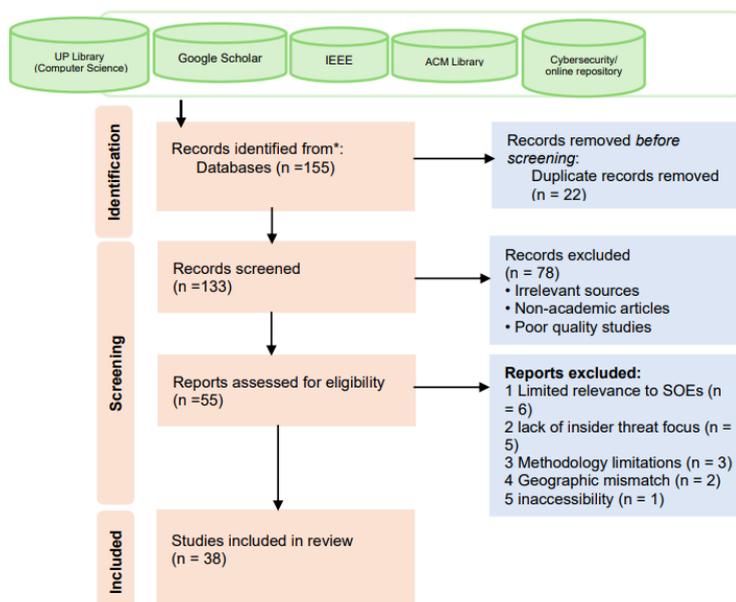

**Figure 2.** PRISMA Methodology

Most of the studies focused on developed economies (n=25), with limited African context research (n=4) available. Table 1 below summarises the studies included per focus area.

| Category | Number % | Focus area |
|---|---|---|
| Insider threat frameworks & detection | 12 (32%) | HMM, behavioural analysis, taxonomies |
| Cybersecurity frameworks & governance | 10 (26%) | AI TRiSM, maturity models, policy |
| Privacy & access control | 6 (16%) | Data protection, authorisation |
| Machine learning applications | 7 (18%) | HMM, RF, explainability (LIME/SHAP) |
| SOE & sector specific | 3 (8%) | Public sector, governance |
| Total | 38 (100%) | |

**Table 1.** Studies Included per Focus Area

Yuan and Wu (2021a) described an insider threat as suspicious activities carried out by individuals within an organisation that negatively impact or cause damage to its valuable assets. Whitelaw et al. (2024) defined insider threats as former or current individuals who have, or previously had, authorised access to sensitive data and other organisational assets. They further emphasised that the evolving nature of insider threats makes them difficult to detect, as insiders already possess legitimate access to information systems. There are different types of insiders, including current employees, former employees, and service providers, who may exploit their access for malicious purposes (Whitelaw et al., 2024). Insider threats are commonly categorised into three groups: insiders who use information technology to conduct malicious attacks, insiders who intentionally steal information, and insiders who commit fraud (Yuan & Wu, 2021). Whitelaw et al. (2024) also described motivational factors that may lead to insider threats, such as employee dissatisfaction, addiction, greed, personal advantage, financial challenges, relationship complications, legal issues, and workplace conflict. Similarly, Allen et al. (2024) explored the motivations of malicious insiders, noting that they can range from financial gain to organisational disgruntlement, personal grievances, or ideological reasons. Understanding these motivational factors is crucial for effectively addressing insider threats. The impact of insider threats on an organisation can be extensive. Sokolowski et al. (2016) demonstrated that insider threats can result in data breaches, intellectual property theft, financial losses, reputational damage, and even legal consequences. Such impacts can severely disrupt organisational operations and compromise integrity, making insider threat management a critical concern for organisations (Mohd & Yunos, 2020). However, the study's limitations include a non-systematic approach to incident identification, the absence of a sustainable linkage between risk and security frameworks, and the lack of performance monitoring mechanisms. Similarly, Hong et al. (2023) proposed an integrated strategy leveraging both manual and automatic feature extraction and employed the ResHybnet model to improve insider threat detection. Despite this, the study revealed several challenges, such as the complexity of mitigating insider threats in daily activities and organisational networks, ineffective dataset utilization, imbalanced classification, the need for deep online learning strategies, and undefined directions for future research. Daubner et al. (2023) established that forensic-ready systems can play a vital role in mitigating insider attacks by enforcing stricter security measures. Daubner et al. (2023) established that forensic-ready systems can play a vital role in mitigating insider attacks by enforcing stricter security measures. Figure 3 illustrates examples of insider threats.

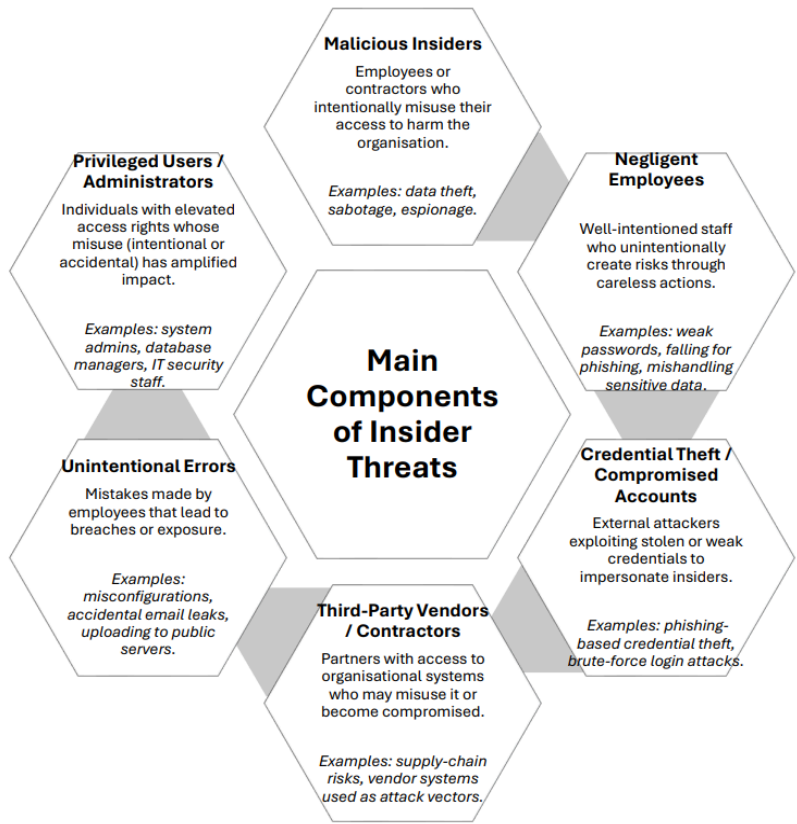

**Figure 3.** Insider Threats Components

Aydin and Sarbas (2022) highlighted that ransomware and hacking (RH) pose severe security threats to industrial control systems (ICSs) and stressed the need for countermeasures. However, the study lacked real-world implementation plans and a detailed analysis of mitigation strategies. Almedires and Almaiah (2021) also evaluated cybersecurity threats, identifying critical areas needing attention, but unresolved challenges remained, such as defining key roles in cybersecurity systems and addressing deployment obstacles. Separhzadeh (2023) suggested using a graph-theoretical simulation approach with directed graphs to model insider threats in organisations, focusing on predicting threat levels without real-world data validation. Nygård and Katsikas (2022) developed an ethical framework for hardware reverse engineering (HRE) to mitigate cyber risks but highlighted the absence of existing protocols and the need for further research. Navdeep et al. (2023) similarly emphasised the importance of ethics in cybersecurity policies, noting the need for a refined ethical framework and in-depth exploration of emerging technologies. Sokolowski et al. (2016) suggested that cultural elements influence insider threat understanding, though empirical evidence did not strongly support this. Existing guides like the CERT Guide offer useful guidelines for insider threat management but have limitations in computational simulation and empirical validation of cultural aspects. Habbal et al. (2024) called for further interdisciplinary research, particularly regarding AI TRiSM frameworks for building trust, while noting the lack of domain-specific investigation for scalability and adaptability. Tyagi (2024) explored integrating blockchain and AI to address cybersecurity threats in IoT and IIoT, showing favourable results but highlighting challenges in scalability, privacy, and ethics. Similarly, Tygia (2020) examined blockchain evolution, identifying challenges and solutions for industries and personal data management, with implications for sustainability, energy efficiency, privacy, and security. Srivastava et al. (2023) stressed integrating blockchain with IoT to address security and privacy threats but noted limitations in storage, scalability, legal issues, resource usage, and smart contract accuracy. Chintala et al. (2024) emphasised the importance of IoT security frameworks, though their proposed framework remains in early development stages. Fernando et al. (2023) investigated cyber risk management integration with suppliers and information security service providers (ISSP), noting improved performance but calling for standardised guidelines and integration strategies. Khaleefah and Al-Mashhadi (2023) reviewed cybersecurity frameworks, highlighting ICT infrastructure availability but weak visibility across subsystems. Shaikh et al. (2023) focused on data protection against insider threats but noted limitations in addressing dynamic changes in e-participation research. Abuasal et al. (2024) emphasised early security requirements in healthcare systems but faced limitations in generalisability due to scarce comparative research.

Nakitende et al. (2023) improved understanding of business fraud and stressed effective leadership and strategy, yet the study lacked balance between surveillance and worker empowerment. Schütz et al. (2023) developed a framework-based information security threat classification to aid employees' understanding of asset-related threats. Pahuja and Kamal (2023) highlighted detecting fraudulent activities in the Ethereum blockchain to enhance investor security; however, limitations included imbalanced datasets, reliance on individual classifiers, and inability to detect linked fraudulent accounts. Milson and Demir (2023) explored data privacy, cyber-attacks, encryption, and regulatory compliance but noted limitations in encryption effectiveness, access controls, and adaptation to emerging technologies. Kafi and Akter (2023) emphasised cybersecurity in accounting data, with limitations in considering human factors and evolving cyber threats. Barnes and Daim (2024) highlighted human elements in healthcare information security but noted potential inconsistencies among expert judgments.

Renaud et al. (2024) stressed addressing insider threat categories through the VISTA taxonomy, calling for a more inclusive approach. Durst and Hinvteregger (2023) examined environmental turbulence impacts on corporate security risk management (CSRM), but research gaps remained in linking turbulence to CSRM. Naman et al. (2024) focused on digital data security enhancements but lacked specific case studies, discussion of obstacles, and financial implications for organisations. Wa Nkongolo (2024) highlighted ransomware transaction detection, with limitations in dataset comprehensiveness and potential data bias. Deep and Aman (2023) explored hidden Markov models (HMM) for cloud intrusion detection, emphasising security measures while noting the need for improved cloud intrusion detection systems. Takale et al. (2024) proposed multi-layered encryption for cloud storage, highlighting ongoing security advancement needs.

Resende and Drummond (2018) demonstrated random forest (RF) effectiveness for threat detection but focused on technical systems rather than organisational assessment. Gaspar et al. (2024) and Hermosilla et al. (2025) applied explainable AI techniques (LIME, SHAP) but did not address practical organisational implementation, highlighting the need for application in state-owned enterprises (SOEs). Gugueoth et al. (2023) analysed blockchain integration for IoT security, noting limitations in privacy, security threat exploration, and evaluation metrics. Ari et al. (2019) highlighted security and privacy in the Cloud of Things (CoT) but emphasised the need for ongoing research, improved protection schemes, and caution regarding weak cryptography. Ophoff et al. (2014) stressed insider threat awareness in information security but lacked focus on governance and compliance. Allen et al. (2024) provided an overview of insider threat mitigation strategies, but their study focused mainly on technical solutions with limited discussion of non-technical approaches. Lee et al. (2019) addressed regulatory challenges in integrating Cyber-Physical Systems and IoT, noting inadequate standards and the need for comprehensive approaches covering trust, privacy, ethics, operational integration, and cybersecurity.

While frameworks such as CRISP-DM, HMM, and AI TRiSM have proven effective individually, limited research exists on their integration for organisational assessment in SOEs. Corrales et al. (2015b) demonstrated CRISP-DM's value in data mining but focused on technical rather than organisational survey data. Rashid et al. (2016) successfully applied HMM for insider threat detection based on user behaviour, yet focused on individuals rather than organisational maturity assessment. This research aims to adapt HMM for enterprise-level security maturity evaluation using survey response patterns. Habbal et al. (2024) demonstrated AI TRiSM's value in AI system risk management but noted cultural limitations and called for systematic collaboration across multiple organisational contexts. This research addresses these gaps by implementing collaborative frameworks in SOE environments.

**Literature Gaps in Insider Threat Management for SOEs**

There are notable gaps in the literature on insider threat management specific to SOEs. Although Ophoff et al. (2014) identified governance perspectives as a critical aspect of insider threat management and highlighted the need for research on governance and compliance, they did not address the unique governance challenges faced by SOEs. Yuan and Wu (2021) provided a thorough analysis of deep learning techniques for insider threat detection but focused primarily on technical detection systems suitable for private-sector applications. Their findings revealed significant challenges, including data variation and imbalanced datasets, which are particularly relevant to SOEs, where security events may occur less frequently but have more severe consequences for national infrastructure. Similarly, Sokolowski et al. (2016) developed agent-based models of insider threats, emphasising the importance of understanding motivational drivers and corporate contexts. However, their framework did not consider the regulatory and accountability requirements that characterise SOE operations, nor was it empirically validated in public-sector settings.

Current literature largely concentrates on technical monitoring and behavioural analysis of individual users, with limited attention to organisational-level assessment through systematic survey methodologies. Barnes and Daim (2024) developed information security maturity models for healthcare organisations using hierarchical decision models and expert panels, demonstrating the value of structured organisational assessments. Nonetheless, their approach was sector-specific and did not account for the unique governance and accountability requirements of SOEs. While Sepehrzadeh (2023) proposed models for assessing insider threats through employee relationship analysis, and Deep and Jain (2023) demonstrated HMM applications for intrusion detection, no existing research has successfully integrated systematic organisational assessment (e.g., CRISP-DM) with behavioural pattern recognition (HMM) tailored to SOE-specific insider threat management.

Resende and Drummond (2018) applied random forest (RF) techniques for technical threat detection; however, their study focused on technical evaluation rather than organisational validation. In contrast, this research utilises RF methodology in combination with HMM modelling to classify SOE security maturity. Gaspar et al. (2024) employed explainable AI techniques, such as LIME and SHAP, for multi-layer perceptron models to enhance interpretability, but their application was limited to network analysis rather than organisational assessment. In this study, explainability techniques are applied to HMM-based SOE security maturity classification, ensuring transparency and actionable insights. Hermosilla et al. (2025) compared the effectiveness of LIME and SHAP in forensic intrusion detection systems, highlighting their technical applications in digital forensics. However, their analysis did not evaluate organisational security maturity, instead focusing on legal and regulatory use of explainable AI. Our research adapts SHAP and LIME to assess and improve security investments and capability development within SOE organisations.

**Methodology**

The research methodology of this study was guided by the research onion model by Saunders, Lewis and Thornhill (2019), utilising a mono-method quantitative approach together with the CRISP-DM Framework. Integrating these two methodologies advances this study by facilitating the development of a novel framework and effective strategies for managing insider threats in SOEs. The first phase of the research involves leveraging the various stages of the research onion (Saunders et al., 2019), in combination with the CRISP-DM phases (Corrales et al., 2015) to construct a comprehensive dataset on insider threats specifically tailored to the SOE context (Figure 5). The second phase incorporates the statistical Machine Learning (ML) model, the Hidden Markov Model (HMM) (Ye & Han, 2022), within the CRISP-DM modelling phase to enable accurate detection of insider threats (Figure 5). These methodologies are particularly suitable for this study as they collectively allow a systematic exploration of insider threats, directly addressing the research questions, tackling the problem statement, and supporting the achievement of the study's objectives.

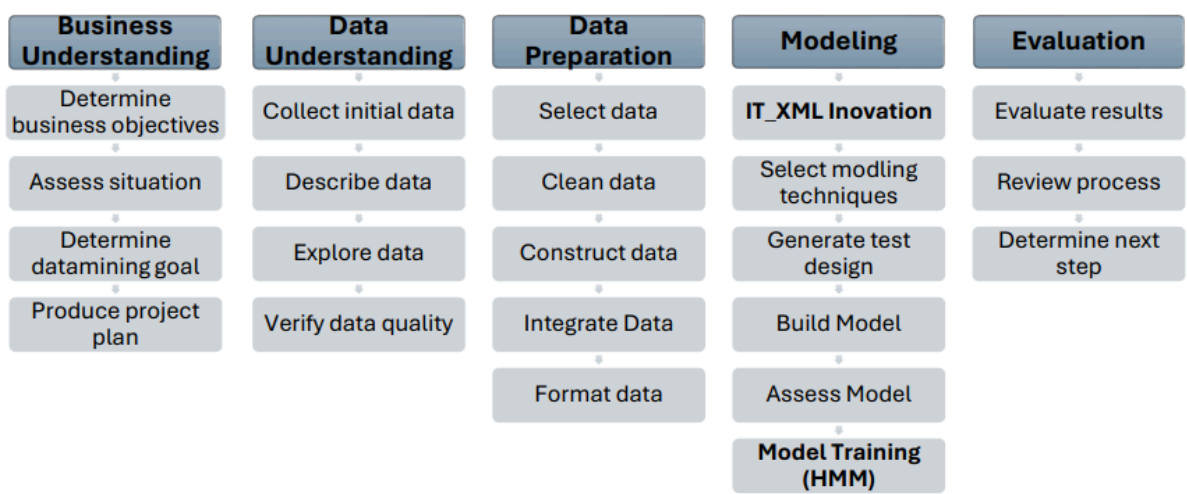

**Figure 5.** CRISP-DM Phases

The proposed IT-XML framework integrating the five phases of CRISP-DM and HMM methodologies to achieve the objectives of this study is shown in Figure 6. Figure 6 illustrates the proposed IT-XML framework, highlighting the current research boundaries and focusing on the first five phases of the CRISP-DM framework: business understanding, data understanding, data preparation, modelling, and evaluation. These phases guided the process of understanding the business context, through which data was collected from the three participating SOEs, ensuring that the study objectives were clearly defined. This approach was essential to identify the types and patterns of insider threats, organise empirical data effectively, and derive meaningful insights. The data understanding phase involved data cleaning and exploration, followed by data preparation, which included recording raw data into numeric format, feature extraction, and integration of data for statistical and HMM analysis. During the modelling phase, the HMM was trained to classify organisational security maturity levels.

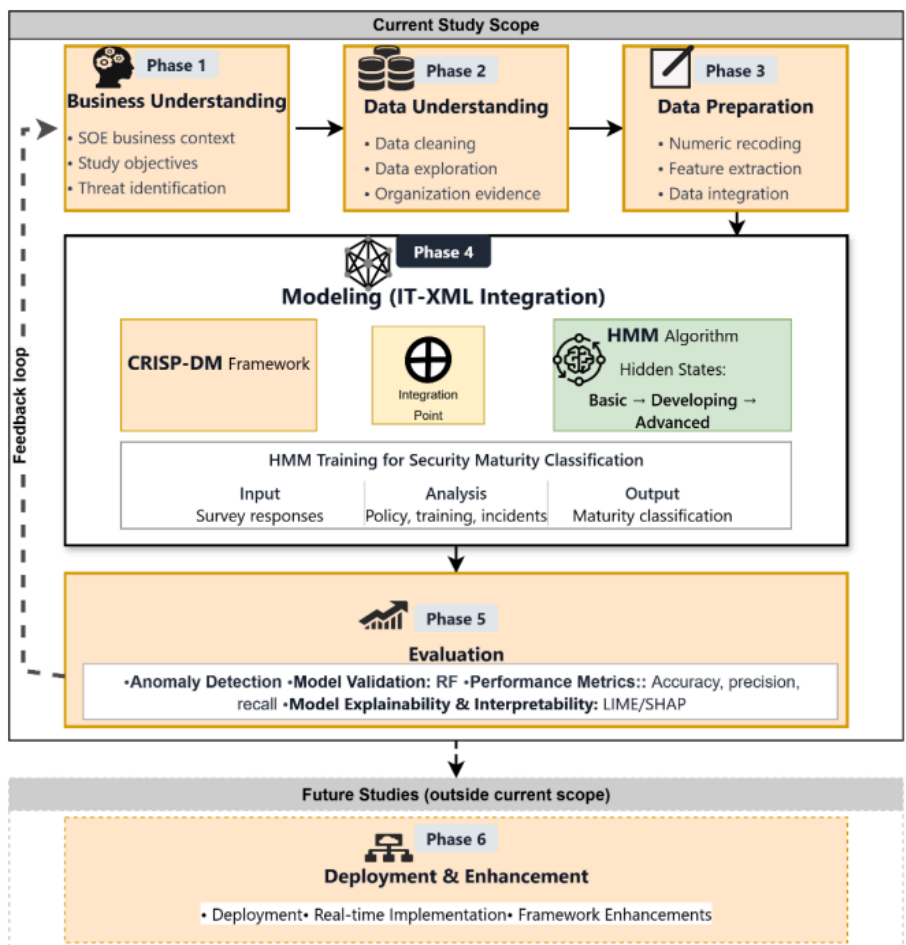

**Figure 6.** The Proposed IT-XML Framework

The evaluation phase focused on anomaly detection, model assessment using random forest (RF), and performance metrics. To enhance transparency and interpretability, explainable AI techniques were integrated: LIME provided local explanations to help stakeholders understand maturity ratings for each organisation, while SHAP determined feature importance, highlighting key factors influencing security maturity classifications. The model also incorporates a continuous improvement loop within the current study scope to enhance framework effectiveness and practicality. The deployment phase is excluded from this study and is reserved for future work, including real-time implementation and framework enhancements. The IT-XML framework demonstrates how organisations can integrate both CRISP-DM and HMM methodologies to manage insider threats effectively. By combining a survey-based approach with advanced Machine Learning (ML) pattern recognition, this study addresses insider threat gaps specific to SOEs. Unlike private-sector organisations, which often rely on technical monitoring, SOEs require methodologies that emphasise regulatory compliance, public accountability, and complex stakeholder environments (Albert & Kabir, 2023). Existing gaps in the literature are also addressed by the IT-XML framework. For instance, Corrales et al. (2015) highlighted the effectiveness of CRISP-DM for data analysis but noted its limited application to organisational datasets. This research extends CRISP-DM to assess organisational data, enabling evaluation of security policies, training effectiveness, and other organisational aspects critical for cybersecurity, as identified by Schütz et al. (2023). Similarly, while Rashid et al. (2016) applied HMM primarily to individual user behaviour patterns for threat detection, this study adapts HMM to classify organisational security maturity states. In this context, the hidden states represent security capability levels (Basic, Developing, Advanced), and the outputs are derived from survey responses regarding policy implementation, training frequency, and incident response capabilities.

**Hidden Markov Model (HMM) Classification Model**

The Hidden Markov Model (HMM) is a probabilistic model used to represent systems that transition between hidden (unobservable) states over time, generating observable outputs at each step (Deep & Jain, 2023). In the context of organisational insider threat management, the HMM classification model is applied to classify the security maturity levels of organisations based on survey data and other behavioural indicators.

**Key Components of the HMM Model in This Study**

1. Hidden States (Security Maturity Levels):

    ○ Represent the underlying organisational security capabilities, which are not directly observable.

    ○ In this study, three levels are defined: Basic, Developing, and Advanced.

2. Observations (Survey Response Patterns):

    ○ Observable outputs derived from survey responses on policy implementation, training frequency, incident response capabilities, and other organisational security practices.

    ○ These observations are probabilistically linked to hidden states, allowing inference of the organisation's maturity level.

3. Transition Probabilities:

- Define the likelihood of moving from one hidden state to another, capturing the dynamics of organisational security development over time.

    - For example, an organisation may progress from Basic to Developing as training and policies improve.

4. Emission Probabilities:

    - Represent the likelihood of observing specific survey response patterns given a particular hidden state.

    - These probabilities allow the model to map observable behaviours to underlying security maturity levels.

5. Initial State Probabilities:

    - Indicate the likelihood of an organisation starting at a particular maturity level before any observations are recorded.

6. Model Training and Classification:

    - The HMM is trained using historical or collected survey data from multiple organisations.

    - The Viterbi algorithm or similar decoding techniques are used to classify each organisation into a hidden state (maturity level) based on observed patterns.

7. Integration with Explainable AI (XAI):

    - Techniques like LIME and SHAP are applied to interpret model outputs.

    - LIME provides local explanations for each organisation's classification, while SHAP identifies the most influential features driving the classification decisions.

Application in the Study
The HMM classification model allows the research to move beyond individual behaviour analysis to organisational-level assessment, providing a systematic approach to measure security maturity and detect potential insider threat vulnerabilities in SOEs. By mapping survey responses to hidden security states, the model generates actionable insights to guide policy, training, and security investment decisions (Rashid, Agrafiotis and Nurse, 2016).

**Random Forest (RF)**
Random Forest is an ensemble Machine Learning (ML) algorithm that constructs multiple decision trees and combines their predictions to improve accuracy and reduce overfitting (Pedregosa et al., 2011). In this study, RF is used to evaluate insider threat detection and organisational security maturity by classifying survey responses and HMM outputs.

**SHAP (SHapley Additive exPlanations)**
SHAP is an explainable AI technique that quantifies the contribution of each feature to the model's prediction (Gaspar et al., 2024). Applied to RF, SHAP highlights the most important factors influencing security maturity classification, helping stakeholders understand which organisational practices drive higher or lower maturity levels.

**LIME (Local Interpretable Model-Agnostic Explanations)**
LIME provides local, instance-level explanations for model predictions (Hermosilla et al., 2025). When used with RF, LIME explains why a specific organisation is classified into a particular security maturity level, offering interpretable insights for decision-makers.

**Linking RF, SHAP, and LIME**
RF serves as the predictive engine for security maturity classification, while SHAP and LIME are interpretability tools that make the RF model transparent. Together, they allow accurate classification of organisational maturity and actionable understanding of the factors influencing insider threat management, ensuring both reliability and explainability of the framework. The IT-XML framework pipeline implementation is detailed in Figure 7 consisting of five sequential stages:

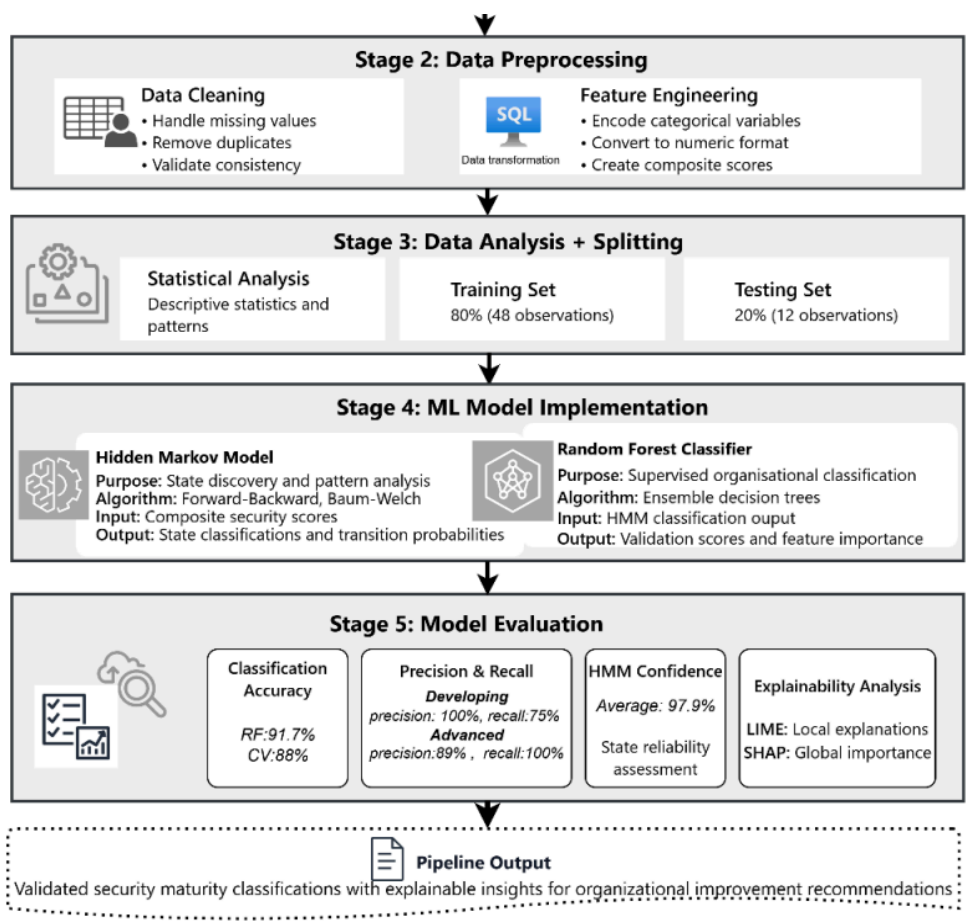

**Figure 7.** IT-XML Framework Implementation Pipeline

**Stage 1: Data Collection**
The survey questions incorporated the AI-TRiSM framework elements by Habbal et al.

(2024) such as model privacy, security application, regulatory compliance and model monitoring. Whereby the questions about data handling procedures, security measures, policy adherence and continuous improvement were incorporated to ensure an inclusive coverage of organisational security dimensions identified as critical for trust and risk management.

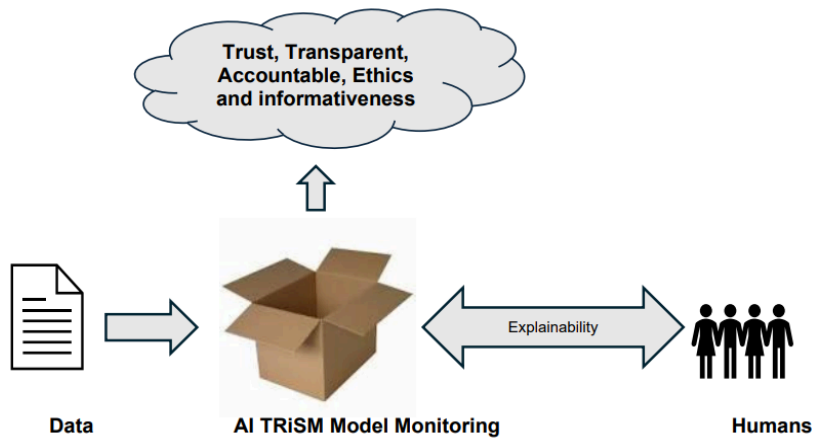

**Figure 8.** AI TRiSM Models Monitoring Functionality (Habbal et al., 2024)

In addition, the survey addressed limitations such as collaboration and cultural consideration highlighted by Nygård & Katsikas (2022) regarding the need for collaboration and cultural considerations. The research utilised the three-phase pre-testing process starting with the subject matter expert review on the questionnaire, employee interviews and a pilot study on how to enhance content validity and cultural appropriateness for the SOE context, to address limitations in the AI TRiSM literature. Data were collected from three SOEs, yielding a total sample of 60 organisational responses. The survey comprised a five-section questionnaire covering insider threat patterns, access control, security measures, policy gaps, and proactive measures. The sample size provided sufficient statistical power for descriptive analysis, correlation testing, and cross-organisational comparisons, while supporting HMM and RF implementation. A power analysis confirmed that 20 participants per SOE were adequate to detect medium effect sizes ($r \geq 0.40$) with 80% power at $\alpha = 0.05$.

**Stage 2: Data Pre-processing**
Data pre-processing involved cleaning, handling missing values, removing invalid responses, validating data consistency, and encoding categorical variables into numeric values using ML techniques. Composite security scores were created for ML input, including security maturity score, threat awareness score, access control effectiveness score, and policy framework score.

**Stage 3: Exploratory Data Analysis**
Descriptive statistics, including mean values and correlation analyses, were performed. Stratified sampling ensured the class distribution was preserved for ML training (80%, 48 observations) and testing (20%, 12 observations). This stage provided insights into data quality, distribution characteristics, and readiness for HMM and RF modelling.

**Stage 4: HMM Implementation**

The HMM (Figure 9) was applied for organisational state discovery using the `hmmlearn` Python library with Baum-Welch parameter estimation (n_components=3, covariance_type="diag", random_state=42) (Saaudi, 2019). Composite security scores from Stage 2 were used to identify hidden organisational states and transition probabilities, which were then mapped to maturity classifications: Basic, Developing, and Advanced (Figure 9).

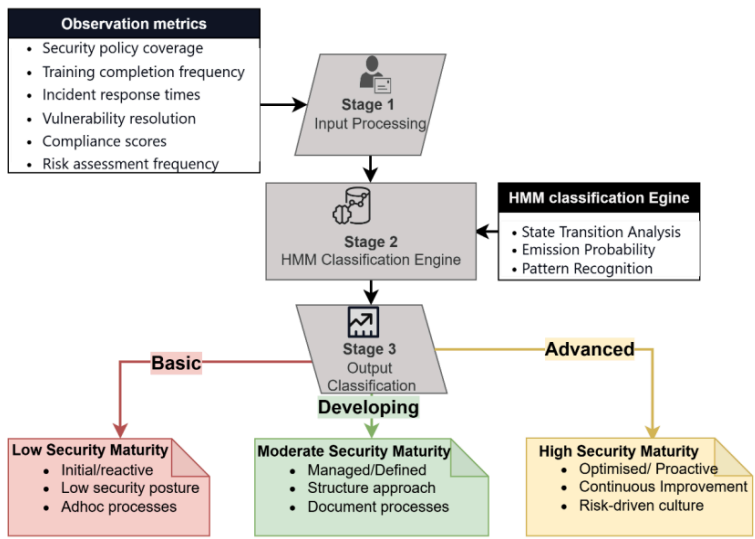

**Figure 9.** HMM Classification Model

**HMM Training Steps**

1. **Data Aggregation.** Observation sequences from all SOEs were vertically stacked using `np.vstack()` to create a unified dataset.

2. **Parameter Initialization.** The model was initialized with three hidden states and diagonal covariance.

3. **Expectation-Maximisation.** The `fit()` method estimated transition probabilities, emission parameters, and initial state distributions.

States were predicted using `hmm_model.predict()`, with probabilities calculated via `hmm_model.predict_proba()` to assess classification confidence. The transition matrix was extracted from `transmat_`, and dominant state classification was determined using `np.bincount().argmax()`. Confidence scores were computed as the mean of maximum probabilities across all time steps. The final classification thresholds were:

- **Basic (0).** Security scores < 2.5 or missing data

- **Developing (1).** Security scores 2.5–3.5

- **Advanced (2).** Security scores > 3.5

The HMM classification model operates in three main stages (Figure 9):

1. **Input Processing.** Organisational metrics are fed into the model as observations. These include security policy coverage, training completion frequency, incident response times, vulnerability resolution, compliance scores, and risk assessment frequency.

2. **State Classification.** The HMM processes the inputs through pattern recognition, computing transition and emission probabilities to classify the current security maturity state of each organisation.

3. **Output Phase.** Organisations are categorised into three maturity levels:

    - **Basic:** Low security posture, limited structured practices.

    - **Developing:** Structured approaches in place but require improvement.

    - **Advanced:** Risk-driven culture with continuous improvement and robust security measures.

This classification approach enables identification of organisational maturity across the participating SOEs, facilitating targeted recommendations for policy enhancement, training, incident response, and other security measures based on the classification outcomes.

**Stage 5: Model Validation and Evaluation**
The labeled data were validated using a Random Forest (RF) classifier with an 80/20 train-test split and k-fold cross-validation. The RF model employed 100 ensemble trees (`n_estimators=100`) with class balancing (`class_weight='balanced'`) and `random_state=42`. Feature vectors included survey responses and HMM-predicted states.

**Performance Metrics**

- **Accuracy.** RF achieved 91.7% classification accuracy with a cross-validation score of 85% (±6.6%).

- **Precision and Recall.** Developing state (100% precision, 75% recall), Advanced state (89% precision, 100% recall).

- HMM state classification confidence averaged 97.5%, confirming robustness.

Table 2 shows all Python libraries, modules, functions, and classes with their types and descriptions utilised to analyse the dataset.

**Explainability**
LIME and SHAP were applied to enhance interpretability. LIME provided local, instance-level explanations, while SHAP quantified feature importance, allowing stakeholders to understand which factors influenced security maturity classification.

| Library/Module/Function | Type | Description | Authors |
|---|---|---|---|
| pandas | Data Analysis | Data manipulation and analysis library for structured data | (McKinney, 2010) |
| numpy | Numerical Computing | Numerical computing operations and array processing | (Harris et al., 2020) |
| scipy | Scientific Computing | Scientific computing library for statistical analysis and hypothesis testing | (Virtanen et al., 2020) |
| scikit-learn | Machine Learning | Comprehensive machine learning library with algorithms like RF and cross-validation | (Pedregosa et al., 2011) |
| hmmlearn | Machine Learning | Hidden Markov Model implementation for sequential data analysis | (Cournapeau et al., 2023) |
| matplotlib | Visualisation | Fundamental plotting library for creating static, interactive visualisations | (Hunter, 2007) |
| matplotlib.pyplot | Visualisation Module | Pyplot interface for matplotlib providing MATLAB-like plotting functionality | (Hunter, 2007) |
| seaborn | Visualisation | Statistical data visualisation library built on matplotlib | (Hunter, 2007) |
| datetime | Built-in Module | Date and time manipulation and formatting | - |
| warnings | Built-in Module | Warning control and filtering system | - |
| os | Built-in Module | Operating system interface for file and directory operations | - |

**Table 2.** Python Libraries

**Outcome**

The IT-XML framework generates evidence-based insider threat management recommendations, producing organisational classifications with confidence measures and actionable improvement directions. By integrating exploratory pattern recognition via HMM with confirmatory validation via RF, this hybrid methodology enables a robust, scalable, and proactive tool for assessing organisational security maturity across SOEs.

**Data Recoding for Survey Responses**

During initial inspection of the survey responses, a critical data type inconsistency was identified. Some responses were stored as text instead of numeric values, which prevented statistical analysis and caused missing values during computations. For example, the survey question: *"How many privacy-related incidents has your organisation documented in the past 12 months?"* offered the options: `"None"`, `"1-2"`, `"3-5"`, `"6-10"`, and `"More than 10"`. Selecting `"None"` resulted in the string `"None"` being stored instead of a numeric 0, making calculations impossible. To resolve this, a systematic recording approach was developed and applied consistently across all affected questions. Using Python within the Anaconda Jupyter Notebook environment, a mapping dictionary was created to convert categorical responses to numeric values. For the privacy incidents question, the recording scheme is presented in Table 3 and Figure 10.

**Ethics**

This study abided by the ethical requirements and procedures specified by the University of Pretoria and the research ethics committee. The questionnaires were administered on a voluntary basis, and the responses were handled in a confidential manner, no respondent's personal information was collected. The response bias analysis indicated that the participation of SOEs was balanced, with 57% of the staff being technical and 43% being non-technical (33.3%) (Figure 11).

| Original Response | Recoded Numeric Value | Rationale |
|---|---|---|
| None | 0 | Represents zero incidents |
| 1-2 | 1.5 | Midpoint of range |
| 3-5 | 4 | Midpoint of range |
| 6-10 | 8 | Midpoint of range |
| More than 10 | 11 | Conservative estimate above threshold |

**Table 3.** Privacy Incidents Response Recording Scheme

```python
incident_mapping = {
    'None': 0,
    '1-2': 1.5,
    '3-5': 4,
    '6-10': 8,
    'More than 10': 11
}
df['privacy_incidents_numeric'] = df['privacy_incidents_column'].map(incident_mapping)
```

**Figure 10.** Python Implementation of Response Recording

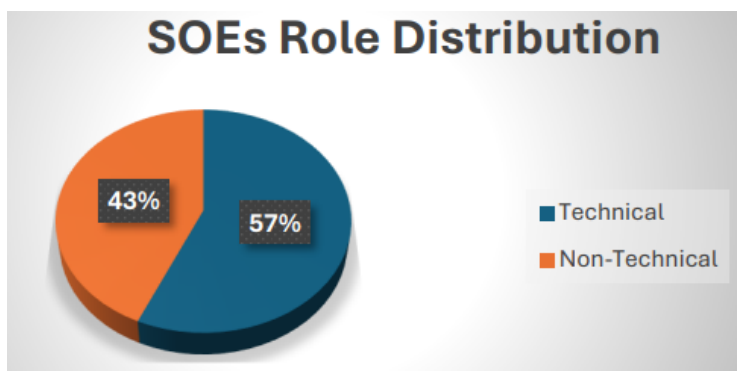

**Figure 11.** SOEs Role Distribution

Further, the researcher first obtained the ethical clearance certificate and organisational approval, which formed the basis of initiating the research. Likewise, the participants were issued with the consent letter stipulating all ethical guidelines, which were handed over to them for signature and questionnaires were only distributed to participants who signed the consent form.

**Data Analysis Results**

Figure 12 demonstrates how HMM analysis applies to the survey-based assessment, moving beyond the original technical log analysis to organisational maturity assessment. The input represents aggregated survey responses from different SOEs, reflecting the types of questions in the questionnaire about insider threats, policy implementation, training frequency, and monitoring capabilities. The state initialisation steps were used for adjusting parameters through the application of the HMM ML algorithms to process and classify the data. Table 4 presents the transition matrix probabilities of organisational improvement or weakening over time (Rashid et al., 2016), while the confusion matrix connects observable survey responses to hidden organisational maturity states (Deep & Jain, 2023). The three algorithms were implemented to perform probability computation, optimal state sequence prediction and parameter optimisation using Forward-Backward (Ye & Han, 2022), Viterbi (Rashid et al., 2016) and Baum-Welch (Deep & Jain, 2023) algorithms respectively. The output provides actionable insights for each SOE, supporting the objective to propose a framework, effective strategies, and recommendations for mitigating insider threats. The composite score of the research dimension across the SOEs together with the descriptive statistics are shown in Table 5. Table 5 results show moderate to high security capabilities across the five dimensions, with Access Control Effectiveness scoring highest (3.91/5.0) and Security Maturity with the lowest score (3.34/5.0), showing the greatest need for improvement. These scores indicate that, despite the fact that all three SOEs have reasonable security foundations, there are distinct opportunities for improvement. The Security Maturity score of 3.34/5.0 indicates that organisations have implemented fundamental security procedures; however, they must allocate additional resources to more advanced threat management capabilities.

This suggests that SOE executives should prioritise budget allocation for security training and system enhancements, as the foundation is in place to build upon. The results in Figure 13 demonstrate that 46.7% of participants reported no privacy incidents in the past 12 months, while 53.3% of participants reported experiencing at least one privacy incident. This means less than half of participants were incident free, while the majority reporting incidents highlights significant privacy incident management challenges in these organisations. This suggests that organisations may have formal incident reporting processes in place, and the presence of incidents requires interventions to improve privacy protection in the organisations (Figure 14). Figure 14 shows that the highest insider threat amongst SOE is information sharing (61.7%), followed by unauthorised access (46.7%), policy violation (45.0%) and data theft (25.0%). The lowest insider threat is system sabotage with 13.3%. Information sharing violations being the highest insider threat type (61.7%) means that employees are either unclear about information handling policies or lack proper training in data classification.

**Machine Learning Results**

Table 5 highlights all the top ten features that were important for HMM analysis. The feature importance analysis demonstrated that vendor breach notification requirements (0.081) is the most important feature for overall cybersecurity effectiveness (Figure 15), followed by regular audit log reviews (0.052) and backup and restore strategy reliability (0.044). The least important feature is readiness to implement new security measures (0.030). A confusion matrix shows the relationship between the predicted values and the actual values in a matrix (Pahuja & Kamal, 2023). The confusion matrix shows the classification accuracy of the ML model performed on the dataset (Figure 16). Figure 16 shows that a total of 12 records were tested, 11 were correctly classified (91.7%). This resulted in 8 advanced records being correctly classified. However, 1 of the developing records out of the 4 were misclassified as advanced. This shows that the model has a strong performance in identifying the organisational security maturity. However, there may be fewer instances where the model misclassifies the maturity level of developing organisations as advanced (Figure 16).

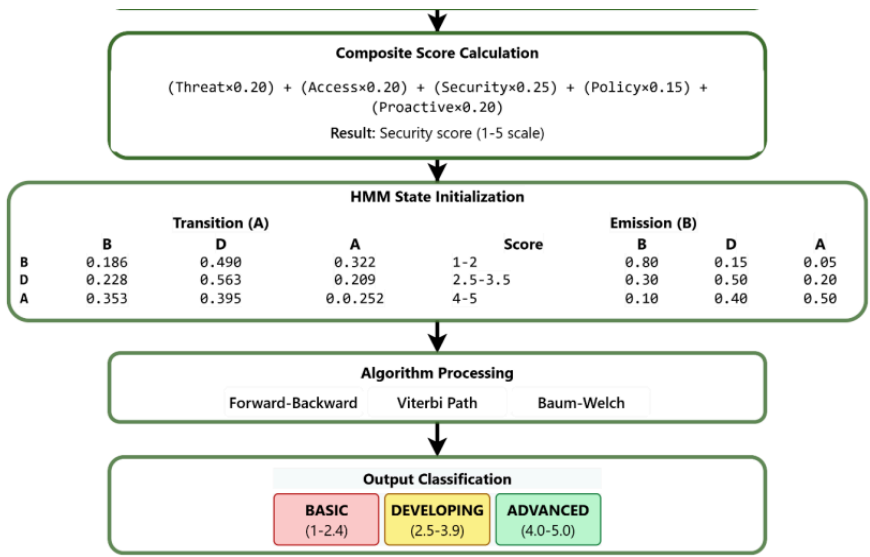

**Figure 12.** HMM for SOE Security Classification

| Actual Maturity Level | Predicted: Basic | Predicted: Developing | Predicted: Advanced |
|---|---|---|---|
| Basic | $TP_1$ | $FN_1 / FP_2$ | $FN_1 / FP_3$ |
| Developing | $FN_2 / FP_1$ | $TP_2$ | $FN_2 / FP_3$ |
| Advanced | $FN_3 / FP_1$ | $FN_3 / FP_2$ | $TP_3$ |

**Table 4.** Transition and Confusion Matrix

| Dimension score | count | mean | std | min | 25% | 50% | 75% | max |
|---|---|---|---|---|---|---|---|---|
| Security maturity score | 60 | 3.34 | 0.55 | 2.38 | 2.94 | 3.19 | 3.77 | 4.43 |
| Threat awareness score | 60 | 3.85 | 0.55 | 2.33 | 3.33 | 4 | 4.33 | 4.67 |
| Access control effectiveness score | 60 | 3.91 | 0.8 | 1.2 | 3.4 | 4 | 4.4 | 5 |
| Policy framework score | 60 | 3.78 | 0.56 | 2.4 | 3.4 | 4 | 4 | 5 |
| Overall security posture score | 60 | 3.7 | 0.44 | 2.87 | 3.4 | 3.68 | 3.97 | 4.58 |

**Table 5.** Descriptive Statistics for Core Assessment Dimensions

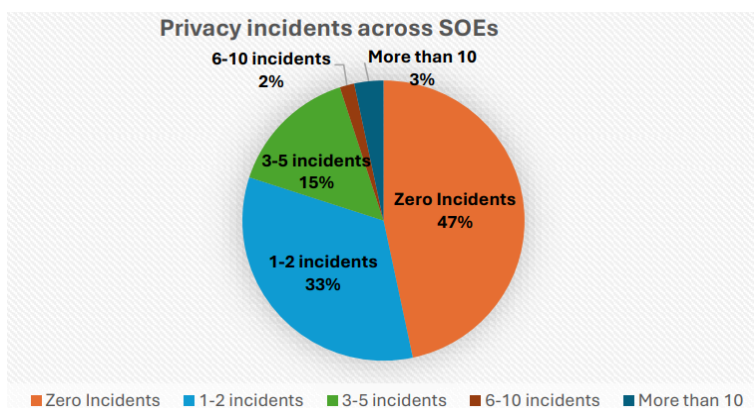

**Figure 13.** Privacy Incidents Across SOEs

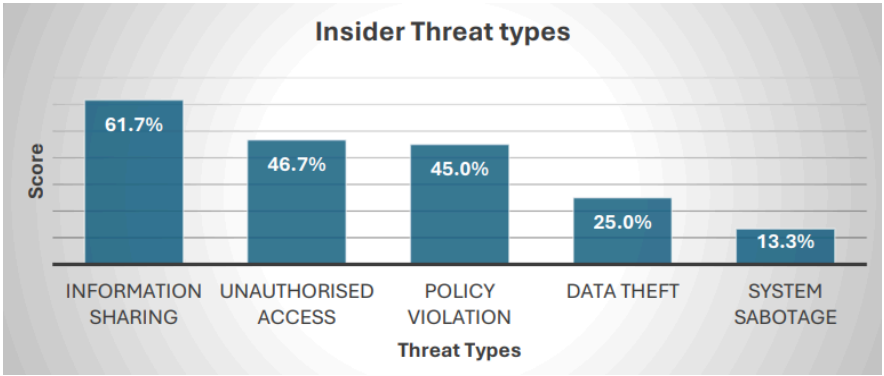

**Figure 14.** Insider Threat Types Across SOEs

| Feature | Importance |
|---|---|
| Rate your organisation's policy framework: Vendor breach notification requirements | 0.080581 |
| Rate your organisation's implementation of these security controls: Regular audit log reviews | 0.0516 |
| Rate your organisation's proactive capabilities: Backup and restore strategy reliability | 0.043874 |
| Rate your organisation's policy framework: Third-party security obligations | 0.04345 |
| Rate the effectiveness of these access controls: Audit log access restrictions | 0.040417 |
| Assess these barriers to policy implementation in your organisation: Management support levels | 0.039739 |
| Rate your organisation's capabilities in the following areas: Identity management of personnel with IT | 0.036218 |
| How frequently are your security policies updated? | 0.030906 |
| Rate your organisation's policy framework: Application security vulnerability assessment | 0.029795 |
| Rate your organisation's readiness to implement new security measures | 0.029482 |

**Table 5.** Feature Importance

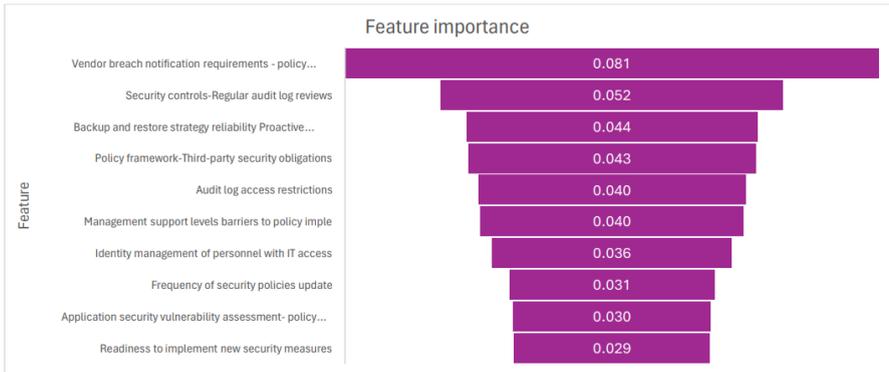

**Figure 15.** Feature Importance Output

Table 6 provides guidance for organisational planning highlighting the precision, recall, F1-score and support across the various organisational levels. Table 6 summarises the output of the model evaluation and validation where organisations were labeled into "Developing" and "Advanced" categories using an automated classification derived from composite security scores. The model misclassified some Developing organisations as Advanced such that SOEs showing improvement may be closer to advancement than initially anticipated. This is encouraging for organisations investing in security improvements since progress is being recognised by the assessment done by the system. However, the high precision for Advanced classification (0.89) means that organisations that are reaching advanced levels have implemented appropriate security measures that could serve as reference for other SOEs. The model showed significant agreement (κ = 0.75), indicating high practical value. Precision discrepancies between the Developing (1.00) and Advanced (0.89) classifications indicate a 11% misclassification rate for advanced organisations facilitating practical implications for organisational planning.

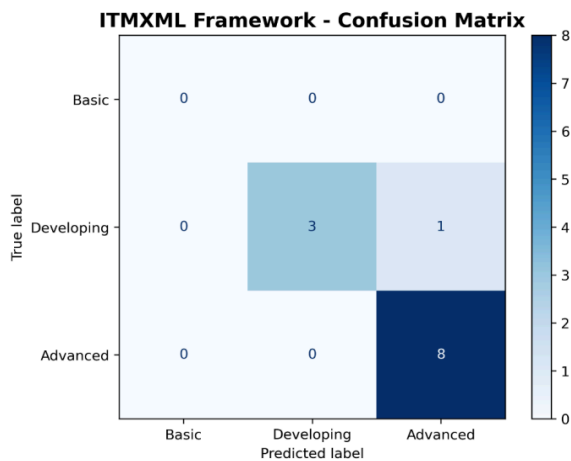

**Figure 16.** Confusion Matrix

|  | Precision | Recall | F1-score | Support |
|---|---|---|---|---|
| **Developing** | 1.00 | 0.75 | 0.86 | 4 |
| **Advanced** | 0.89 | 1.00 | 0.94 | 8 |
| **accuracy** |  |  | 0.92 | 12 |
| **macro avg** | 0.94 | 0.88 | 0.90 | 12 |

**Table 6.** ML Classification Report

After the final computation, the labeled dataset contained: 4 developing organisations (33.3% of dataset) and 8 advanced organisations (66.7% of dataset). Table 7 shows there was no categorisation for the basic state which resulted in only two classes in the final analysis. This means that no organisations in the dataset were below the 2.5 score.

| Actual | Predicted | Actual Label | Predicted Label |
|---|---|---|---|
| 1 | 1 | Developing | Developing |
| 2 | 2 | Advanced | Advanced |
| 2 | 2 | Advanced | Advanced |
| 2 | 2 | Advanced | Advanced |
| 2 | 2 | Advanced | Advanced |
| 1 | 1 | Developing | Developing |
| 2 | 2 | Advanced | Advanced |
| 1 | 1 | Developing | Developing |
| 2 | 2 | Advanced | Advanced |
| 2 | 2 | Advanced | Advanced |
| 2 | 2 | Advanced | Advanced |
| 1 | 2 | Developing | Advanced |

**Table 7.** Labeled Dataset

The HMM analysis classified each SOE into security maturity states based on composite scores across five dimensions. Table 8 illustrates the organisational maturity of the three SOEs and they all dominated in the Developing State. This reveals that all SOEs have moderate capabilities and will need improvements to transition to the advanced stage.

| Company Code | Dominant State | Confidence | Basic Count | Developing Count | Advanced Count |
|---|---|---|---|---|---|
| 0 | Developing | 0.982 | 6 | 10 | 4 |
| 1 | Developing | 0.974 | 4 | 10 | 6 |
| 2 | Developing | 0.980 | 5 | 11 | 4 |

**Table 8.** HMM State Classification for Each SOE

**HMM Transitions Probabilities**

The HMM model identified possible transitions for organisational improvement between the basic, developing and advanced state (Figure 17). The HMM transition matrix in Figure 17 was generated through HMM computation trained on composite security scores from all SOEs. The model identified three hidden states (Basic, Developing, Advanced) and learned the transition probabilities between these maturity levels. The results show that there is a high persistence (56.3%) in the developing state while basic (18.8%) and advanced (25.2%) have low persistence rates. However, the analysis in the first row suggests that organisations have a 18.8% probability of remaining in the basic state, 49% of moving to developing and 32% of jumping to the advanced state. In the second row, the organisation has a 22.8% probability of falling to basic, 56.3% of staying in the developing stage and 20.9% chance of moving to advanced state. Thirdly, the organisation has a 35.3% probability of falling back to basic, 39.5% of dropping to developing and 25.2% chances of staying in the advanced state. The developing state proved to be the most stable with 56.3%. However, with targeted interventions in place, all SOEs have a 20.9% chance of advancing to Advanced state. The 56.3% probability of staying in Developing state indicates that without targeted interventions, SOEs will likely maintain current security levels. However, the 20.9% chance of advancing to Advanced is achievable with improvement initiatives. The lowest 22.8% chance of falling to Basic level emphasises the importance of sustained security investment for instance cost cutting security budgets may result in capability dropping. For yearly planning, SOEs should view security improvement as requiring consistent, sustained investment rather than one-time initiatives. The 56.3% probability of remaining in Developing state representing a high stability effect (w = 0.62 using Cohen's conventions for categorical data). The 20.9% advancement probability indicates that targeted interventions have moderate potential impact, with approximately 1 in 5 organisations advancing with sustained effort. The proposed IT-XML framework demonstrated high performance with 88.9% accuracy and ensured complete data coverage across all the three SOEs analysed (Figure 18). Figure 18 demonstrated the robust performance of about 91.7% accuracy and 85% cross validation score, while analysing all 63 security features across the three SOEs.

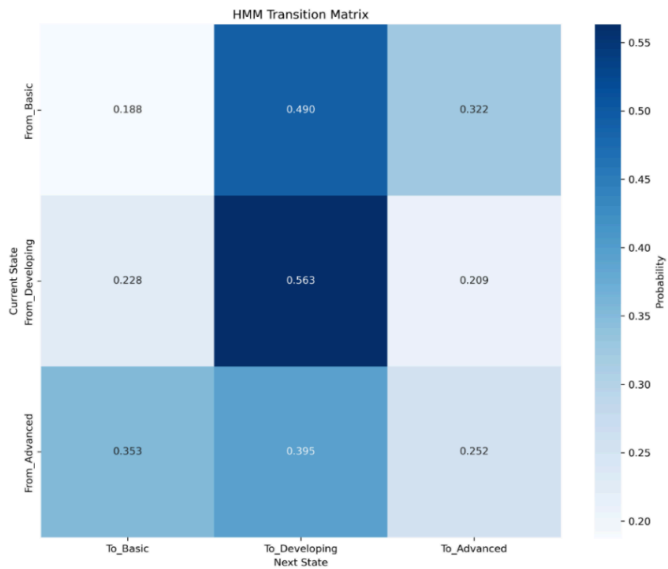

**Figure 17.** HMM Transition Matrix

The 91.7% accuracy rate means that the IT-XML framework provides reliable organisational assessment that can guide investment decisions with confidence. The complete data coverage across all three SOEs demonstrates that the framework works regardless of SOE size or sector. Most importantly, the ability to analyse 63 security features means that security investments can be prioritised based on evidence rather than instincts or vendor recommendations. This is particularly valuable for SOEs that must justify expenditures to the governing bodies.

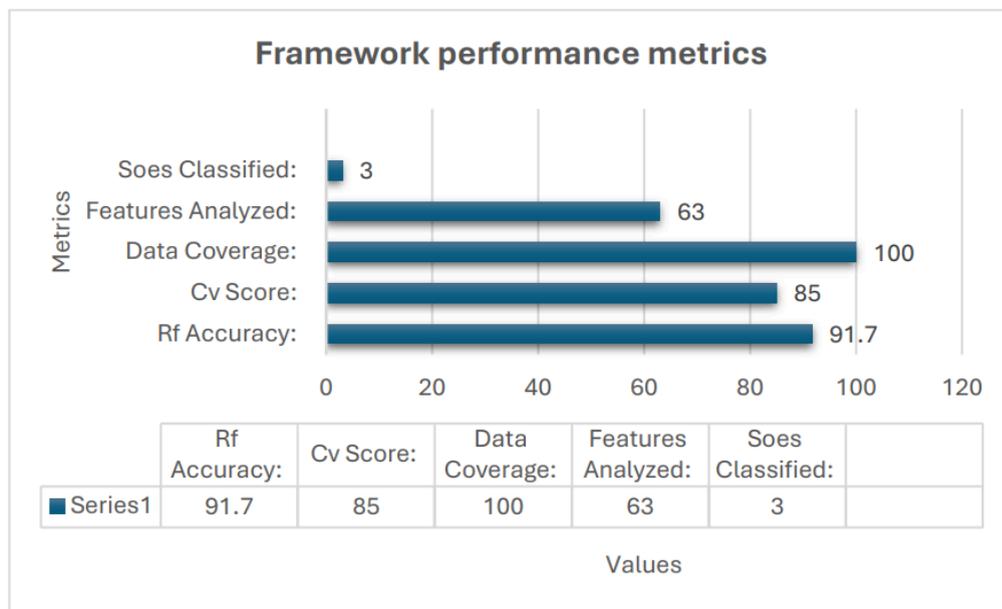

**Figure 18.** Framework Performance Metrics

**Model Explainability Analysis**

RF feature importance rankings were validated through SHAP analysis, producing a correlation coefficient of r = 0.942 (p < 0.001), confirming 94.2% consistency between the two methods (Table 6).

| Rank | RF Importance | Feature | Security Domain |
|---|---|---|---|
| 1 | 0.0806 | Vendor breach notification requirements | Policy Framework |
| 2 | 0.0516 | Regular audit log reviews | Security Controls |
| 3 | 0.0439 | Backup and restore strategy reliability | Proactive Capabilities |
| 4 | 0.0435 | Third-party security obligations | Policy Framework |
| 5 | 0.0404 | Audit log access restrictions | Access Controls |
| 6 | 0.0397 | Management support levels | Policy Implementation |
| 7 | 0.0362 | Identity management of personnel with IT access | Security Audit & Governance |
| 8 | 0.0309 | Security policy update frequency | Policy Framework |
| 9 | 0.0298 | Application security vulnerability assessment | Policy Framework |
| 10 | 0.0295 | Readiness to implement new security measures | Proactive Capabilities |

**Table 6.** RF Features Ranking

This strong relationship demonstrates that the identified features represent strong predictors of security maturity. Table 6 demonstrates that policy framework components dominated the top rankings of 1 and 4 respectively, emphasising the importance of regulatory compliance and vendor management. The feature importance ranking is indicating that vendor management, regular audit log reviews, backup and restore strategy, third party security and audit log restrictions are fundamental predictors of security maturity. Operational security controls such as audit log reviews and access restrictions appear twice in the top 5, highlighting the importance of monitoring capabilities while proactive capabilities of backup and restore ranked third, highlighting the significance of business continuity planning. Features such as management support for policy implementation, identity management, application security and proactive capabilities to implement new security measures were also among the most important predictors. The analysis of the top five features only out of 63 security dimensions shows a combined importance score of 0.270 (27%). This concentration demonstrates that strategic investments on vendor breach notification requirements, regular audit log reviews, backup and restore reliability, third-party security obligations, and audit log access restrictions can deliver significant security maturity improvements without targeted interventions on all other organizational security domains. SOEs with resource constraints will benefit from focused investment strategies over spread resource allocation methods. Figure 19 shows SHAP global feature importance rankings for the top ten security factors. Vendor breach notification requirements demonstrated the highest importance (0.0482), followed by regular audit log reviews (0.0315) and identity management capabilities (0.0277). The visualization confirms that policy framework and operational monitoring components dominate security maturity predictions. Figure 20 shows SHAP feature impact distribution across individual predictions. Each point represents a test sample, with colour indicating feature value whereby blue indicates low, and red is high. The horizontal spread of the dots demonstrates that most features show consistent directional impact, with higher feature values (red dots) generally associated with positive SHAP values aligning with the expected relationships between security practices and maturity classifications.

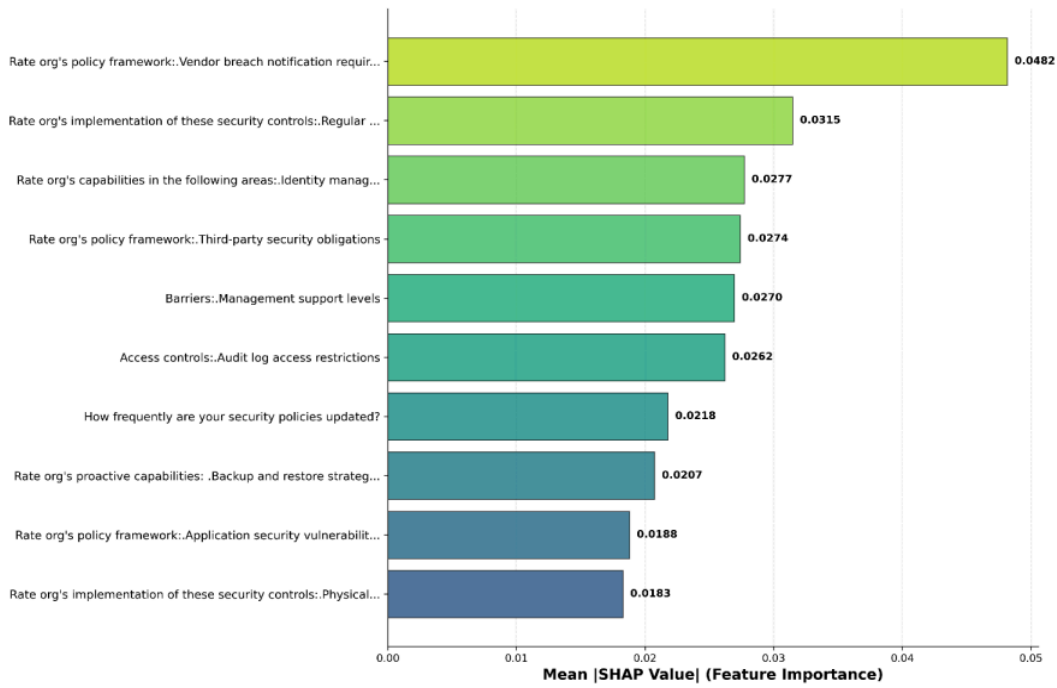

**Figure 19.** SHAP Feature Importance Positive Classification

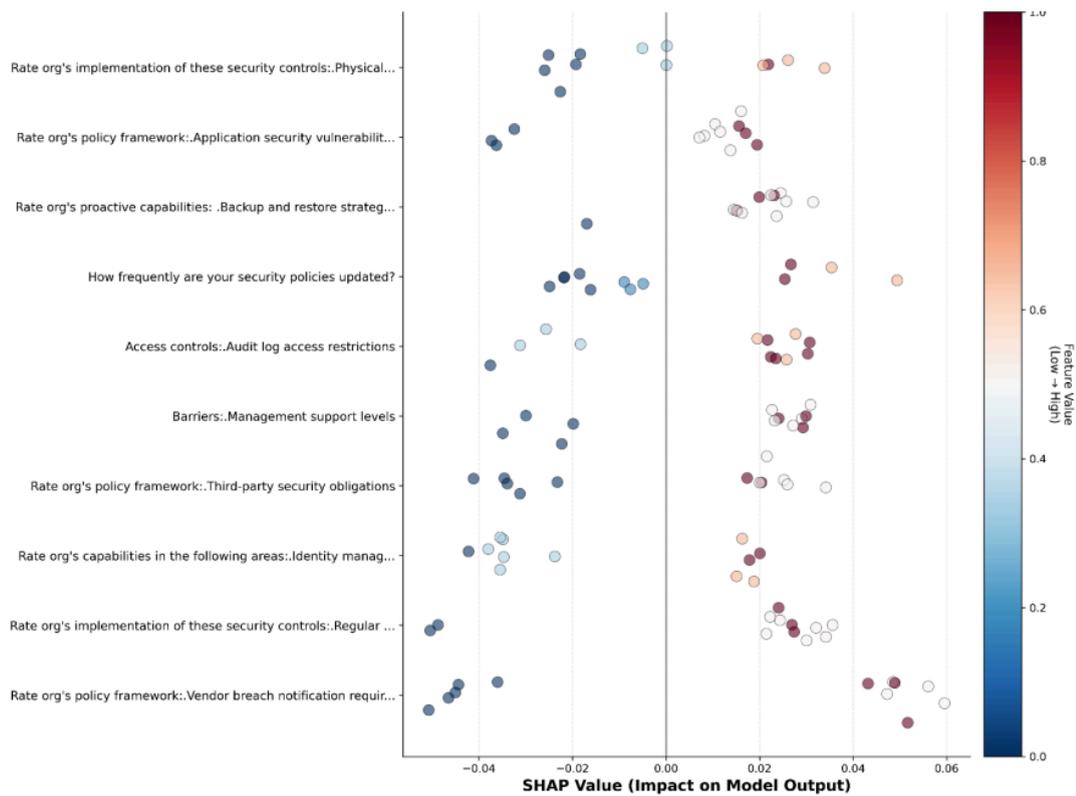

**Figure 20.** SHAP Feature Impact Positive and Negative Class Classification

**Specific Local Explanation Using LIME**

Figure 21 shows the LIME explanations revealing varying patterns of security maturity and implementation challenges across the participating organisations. The first organisation demonstrates significant weaknesses in translating policy frameworks into operational security measures. The strongest negative contributors to its classification include policy framework issues and limited management support for implementation, followed by deficiencies in physical and logical security controls and infrequent security updates. These findings indicate systemic barriers in policy enforcement and insufficient integration of security measures across the organisation's operations.

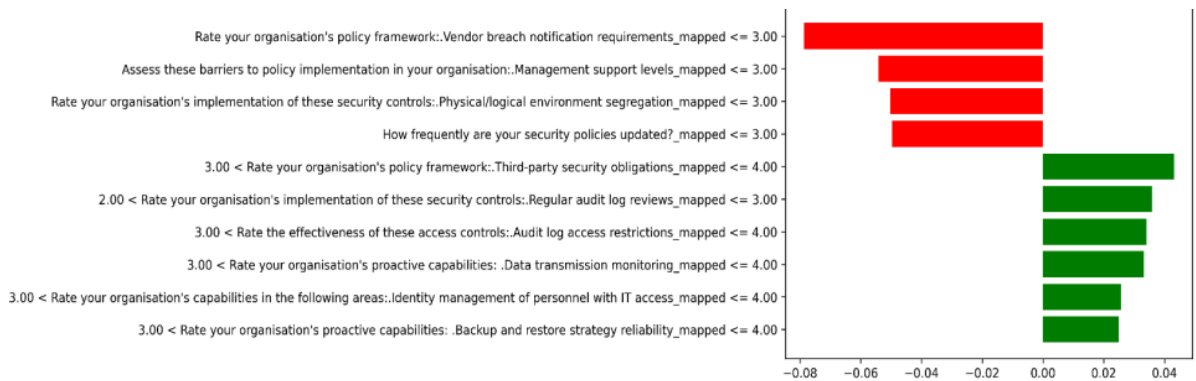

(a) First SOE

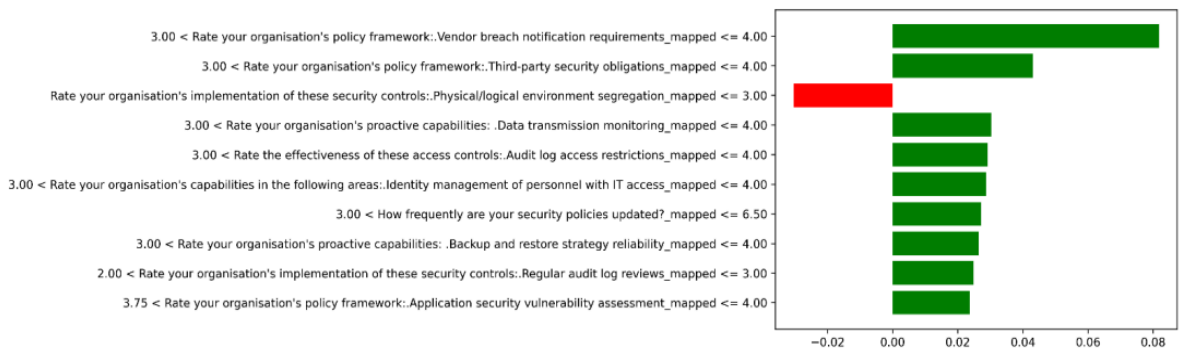

(b) Second SOE

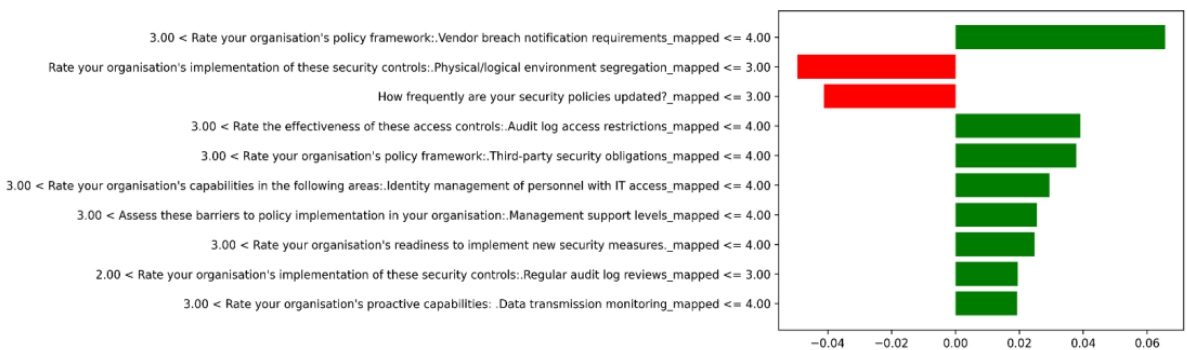

(c) Third SOE

**Figure 21.** Overall LIME Results

The second organisation presents a more balanced security posture, showing notable strengths in policy framework implementation and third-party security measures (Figure 21). Despite minor challenges in implementing certain security controls, its overall performance reflects a mature and stable security environment. This organisation exhibits the most well-rounded security maturity profile among the three, suggesting stronger policy enforcement and oversight mechanisms. The third organisation demonstrates solid policy development practices but faces persistent implementation challenges. Although policy frameworks positively contribute to its classification, issues such as inconsistent security control application and irregular policy updates negatively affect its performance (Figure 21). This suggests that while the organisation is strong in strategic policy formulation, it requires improvement in executing and maintaining operational security processes.

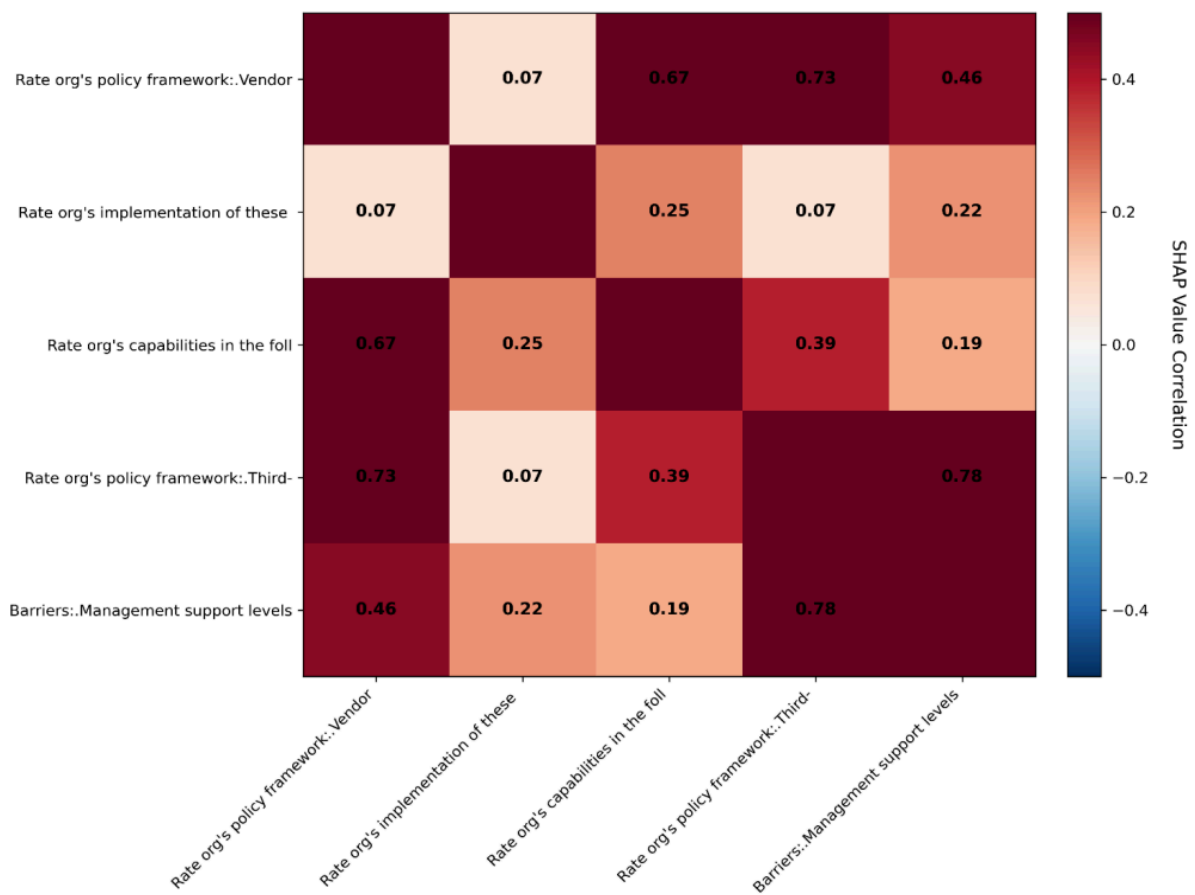

**Figure 22.** SHAP Summary

Figure 22 demonstrates SHAP value correlations among the top 10 five features. The correlation values are between 0.07 to 0.78 showing the relationship between feature pairs. The strongest correlation is between third-party security obligations and management support levels at 0.78. However, most of the pair correlations were below 0.5, indicating that most features do not depend on each other. This is beneficial for SOEs because they can work on security improvements one area at a time without needing to implement everything at once instead of trying to fix everything simultaneously, organizations can focus their resources on individual security domains and still see incremental improvements.

**Discussion**

The explainability analysis validates the reliability and robustness of the IT-XML framework through both performance and interpretability measures. The Random Forest (RF) model achieved a high classification accuracy of 91.7% and a cross-validation performance of 85.0%, indicating strong and consistent predictive capability. A strong correlation (r = 0.942) between SHAP feature importance and RF outputs further confirms that both methods consistently identified the same critical factors influencing organisational security maturity. LIME local explanations provided deeper insights into organisation-specific patterns across all 60 survey responses. One organisation exhibited implementation challenges with a negative average contribution, another demonstrated a balanced and well-structured security posture with positive contributions, while the third showed moderate security maturity with room for improvement. These local-level insights help pinpoint specific strengths to sustain and weaknesses that require targeted interventions. Overall, the integration of SHAP (global interpretability) and LIME (local interpretability) highlights that policy frameworks and operational monitoring are reliable, recurring predictors of security maturity across the studied organisations. The alignment between these explainable AI methods ensures that the derived insights reflect genuine organisational priorities, thereby offering trustworthy, data-driven guidance for future security strategy and investment decisions.

**Limitations**

Despite the promising outcomes of the explainability analysis, several limitations must be acknowledged. First, the dataset size was relatively small, limited to 60 survey responses from a few organisations, which may affect the generalisability of the model to broader institutional contexts. Second, the survey relied on self-reported data, introducing the potential for response bias or subjectivity in assessing security practices. Additionally, while the Random Forest and HMM models performed well, their effectiveness depends on the quality and consistency of the input variables—any inconsistencies or missing data could influence the accuracy of classification results. Finally, the use of SHAP and LIME provided valuable interpretability, but these methods can be computationally intensive and may not fully capture complex interdependencies between features in high-dimensional security datasets.

**Future Directions**

Future research should aim to expand the dataset to include more organisations across different sectors to enhance the generalisability and robustness of the IT-XML framework. Incorporating real-time monitoring data, such as system logs or network traffic patterns, can provide a richer and more objective measure of security maturity. Moreover, hybrid explainability models combining SHAP, LIME, and causal inference methods could offer deeper insights into cause-effect relationships influencing organisational security. Exploring the integration of deep learning models with enhanced interpretability could also strengthen predictive accuracy while maintaining transparency. Finally, longitudinal studies could be conducted to track how targeted interventions—guided by model insights—improve organisational security over time.

**Conclusion**

In conclusion, the IT-XML framework, supported by explainable AI techniques such as Random Forest, SHAP, and LIME, effectively classified and explained security maturity levels among the analysed organisations. The combination of high model performance and interpretability confirms that the approach provides actionable insights into organisational strengths and weaknesses. Although limitations exist regarding data scope and model assumptions, the framework establishes a solid foundation for evidence-based decision-making in cybersecurity management. Future enhancements focused on scalability, automation, and richer datasets will further refine the model's accuracy and strengthen its potential as a strategic tool for improving information security governance across diverse organisations.